# Ultra-Broadband plug-and-play photonic circuit packaging with sub-dB loss

Erik Jung [a], Helge Gehring [b], Frank Brückerhoff-Plückelmann [a], Linus Krämer [a], Clara Vazquez-Martel [c], Eva Blasco [c], Wolfram Pernice [a,b]

[a] Heidelberg University, Kirchhoff Institute for Physics, Im Neuenheimer Feld 227, Heidelberg, Germany
[b] Institute of Physics and Center for Nanotechnology, University of Münster, Münster D-48149, Germany
[c] Heidelberg University, Institute for Molecular Systems Engineering and Advanced Materials, Im Neuenheimer Feld 225, Heidelberg, Germany

**Abstract**. Photonic Integrated Circuits (PICs) offer transformative potential for computing systems, enabling high-bandwidth neuromorphic processors and facilitating low decoherence quantum computing on a chip scale platform. However, the development of robust and scalable optical packaging solutions remains a major challenge. Efficient fiber-to-chip coupling is essential for minimizing loss and enabling high optical bandwidth, key requirements for photonic computing.

Here, we introduce a novel plug-and-play solution for fiber-to-PIC connections using female multi-fiber termination push-on cables and additively fabricate the alignment counterpart on the circuit via two-photon polymerization. We develop 3D out-of-plane couplers, offering a peak transmission of −0.41 dB and broadband performance with losses below 0.55 dB across the 1500–1600 nm range. Integration of the couplers with the plug-and-play solution adds in average only 0.37±0.12 dB of loss, setting a record in passive out-of-plane packaging solution with losses of 0.78 dB, as well as -0.5 dB bandwidth greater than 100 nm.

We characterize the reproducibility of this out-of-plane packaging solution in terms of losses, as well as bandwidth by interfacing a 17-port photonic circuit for incoherent photonic computing. The high bandwidth of the packaging is crucial to couple the full 100 nm bandwidth spectrum of a superluminescent light emitting diode, which consequently enables low noise computing at 17.6 GBaud.

Our concept enables multiport, passive and reconfigurable photonic integrated circuit packaging providing reliability and versatility driving photonic packaging towards the scalability and robustness of electric chip packaging.

*First Author, E-mail: erik.jung@kip.uni-heidelberg.de

# 1 Introduction

With its high data throughput, low latency and reduced energy consumption, Photonic Integrated Circuits (PICs) offers a compact and efficient platform for technological advancements. PICs hold immense potential in fields such as sensing [1], [2], [3], quantum applications [4], [5], [6], optical communication [7], [8] and neuromorphic computing [9], [10], [11].

The success of PICs relies on low-loss, broadband and robust optical packaging compatible with electronic standards to ensure optimal performance, a challenge across various material platforms [12], [13].

Efficient coupling requires compensating for the mode field diameter difference between on-chip waveguides and optical fibers used to address the PICs. Fiber-to-chip coupling is achieved through two main approaches [14], [15]: in-plane and out-of-plane coupling.

For in-plane coupling, the chip interfaces with the fiber through a cleaved facet. Spot size adaptation with microlenses [16], [17], spot size converters [18] [19], or tapered fibers [20] enhance coupling efficiency, but the requirement of edge waveguide routing restricts the number of addressable devices, reducing design flexibility and wafer-scale testing capabilities.

Out-of-plane coupling resolves this limitation by redirecting the beam out of the chip's plane. Typically this is done via grating couplers fabricated within the photonics layer capable to achieve a $-3$dB transmission due to symmetric diffraction [21], [22]. Apodization combined with advanced fabrication like backside mirrors improve efficiency up to -0.55 dB [22], [23], [24]. However, their interference based working principle limits bandwidth. Out-of-plane polymer couplers combine the high bandwidth and low losses with wafer-scale technology and can be produced by Two-Photon Polymerization (TPP) [25], [26], [27], [28], [29], [30].

Conventional packaging techniques, such as actively aligned fiber arrays [31], [32] or freeform optical wirebonds [16], [33] are non-reconfigurable and lack scalability [34]. So far, passive and removable plug-and-play solutions are limited to grating couplers combined with 3D-printed guiding structures [35] [36], 3D printed lenses relaxing sensitivity to alignment tolerances [37], [38], or rely on aligned interposer chips [39].

In this paper, we present a self-aligning, multiport, low loss, high bandwidth plug-and-play solution. Combining a female Multi-Fiber termination Push-on (MTP) cable and micron-precise

TPP-printed alignment counterparts on the PIC surface, light is coupled via 3D out-of-plane couplers based on Total Internal Reflection (TIR) couplers. By comparing Finite Element Frequency Domain (FEFD) simulations with experimental results, we optimize the TIR couplers. The ultra-broadband, low-loss couplers are passively packaged, resulting in a loss of 0.41 dB for the coupler and 0.37 dB for the packaging. We demonstrate the excellent reproducibility of this macroscopic packaging solution comparable to a USB plug for PICs by addressing an optical processor for matrix vector multiplication (MVM) with 17 ports.

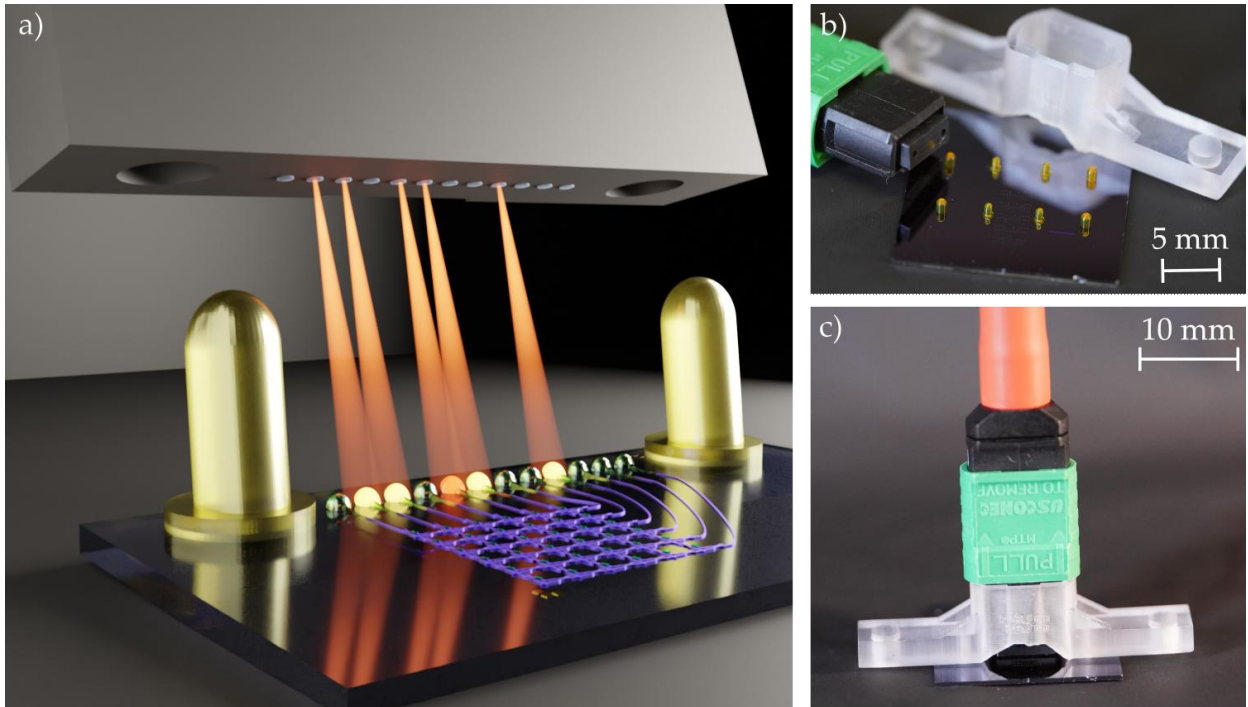

**Fig. 1** *Plug-and-play concept:* (a) Illustration of the plug-and-play solution: The PIC, shown in blue, is addressable using total TIR couplers. The couplers are interfaced using an 8 degree polished female MTP cable, a fiber array with directly integrated alignment pin holes. Alignment between the PIC and the fiber array is achieved via TPP-printed alignment pins, directly fabricated onto the chip's surface. (b) A PIC featuring two sets of pins for a plug-and-play interface, designed to connect 12×2 input ports via an MTP cable. Mechanical stability is ensured by a DLP-printed pre-alignment structure, which both supports the MTP cable and bears its weight.
(c) The fully assembled system, precisely aligning the MTP cable ports with the PIC.

## 2 Results

A conceptual image of the plug-and-play solution is shown in Fig. 1(a). The on-chip nanophotonic waveguides (blue) forming a photonic matrix are interfaced with polymer out-of-plane couplers (green). These couplers redirect the light beam out of the chip's plane and are addressed with the help of an MTP cable featuring an 8° angle polished facet to reduce backscattering. The MTP cable plugs into a counter piece fabricated by direct laser writing on the chip's surface allowing for addressing of the individual PIC ports. Using a direct laser-printed connector, shown in Fig. 1(b), the MTP cable is pre-aligned with the chip under the respective 8° angle. This pre-alignment structure also bears the weight of the MTP cable, ensuring the mechanical stability of the connection. The fully assembled structure is shown in Fig. 1(c). The ends of the connector can be screwed into a PCB further increasing the stability of the connection. Recesses in the pre-alignment structure allow it to remain fully compatible with conventional electrical packaging standards.

*2.1 Coupler Optimization*

The transmission from fiber to PIC is determined by the transmission of polymer out of plane couplers. For that reason, based on FEFD simulations using COMSOL Multiphysics, we optimize the transmission of the polymer out of plane couplers. We perform this optimization for the thin film $Si_3N_4$ platform, but the results can be transferred to different material platforms. To interface with PICs, optical elements are accessed using millimeter-sized fiber arrays, which need to be positioned with micron-level precision relative to the chip. To preserve the integrity of submicron optical elements, mechanical spacing between the fiber array and coupler is crucial. Direct contact could damage the coupler or compromise its optical properties; therefore, a horizontal free-space section is employed to maintain functionality.

The coupler's efficiency relies on replicating the natural divergence of a beam emerging from a Single Mode Fiber (SMF). Therefore, we calculate the divergence of the beam using the Gaussian beam divergence formula for a SMF-28 fiber [40].

$$w(l_f) = w_s = w_f \cdot \sqrt{1 + \frac{l_f}{z_R}} \text{ with } z_R = \frac{\pi \cdot w_f^2}{\lambda}, \qquad (1)$$

where $w_s$, $w_f$ and $z_r$ are the mode field diameter at the focus point, the mode field diameter at the coupler's surface and the Rayleigh length, respectively.

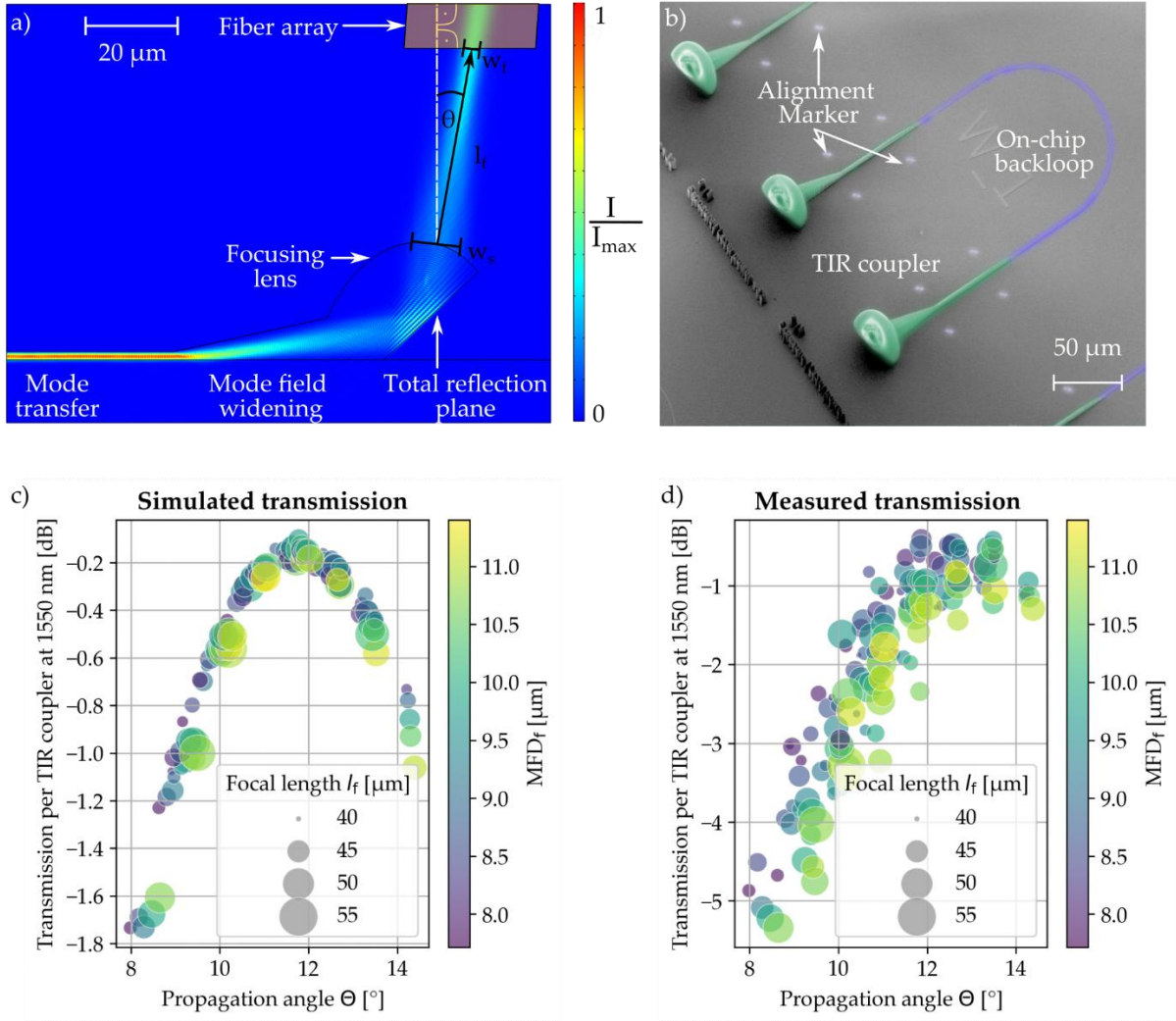

**Fig. 2** *Coupler Optimization:* (a) FEFD simulation result displaying normalized power propagation within the coupler structure, highlighting parameters affecting coupling efficiency. (b) SEM image of an U-shaped $Si_3N_4$-waveguide backloop interfaced with TIR couplers used to determine the coupling efficiency per coupler. Comparison of simulated (c) and experimentally measured transmission (d) as functions of the beam's propagation angle, MFD at the coupler focus, and focal length. Each data point represents the transmission at 1550 nm of one coupler geometry following positional optimization.

A 2D FEFD analysis of the coupler's cross-section is presented in Fig. 2(a), illustrating the normalized power propagation within the structure. The TIR coupler features four distinct sections:

In the *mode transfer* section, optical power shifts from the single-mode $Si_3N_4$ on-chip waveguide to a single-mode polymer waveguide. To ensure efficient coupling, the $Si_3N_4$ waveguide is tapered down, enabling an adiabatic transition from a $Si_3N_4$-like mode to a polymer waveguide mode, with experimentally demonstrated losses of 0.12 dB per transition (see Supplement A1). To match the MFD of the polymer waveguide with that of the SMF, the *mode field widening* section uses a non-adiabatic taper to expand the MFD. This free-space like propagation accelerates the MFD expansion and suppresses reflections from substrate/air interfaces. The beam subsequently reflects off a *total reflection plane*. This plane is angled precisely to redirect the beam out of the chip plane to an 8° angle-polished fiber array, which collects the light of the beam at an angle of 11.8° in air.

The output ellipsoidal *focusing lens* of the coupler is designed to align the beam's mode field diameter (MFD) with that of the Corning SMF-28 fiber (~10.4 μm at 1550 nm) at the focal point. By adjusting the in-plane radius, the focusing strength is controlled, while the out-of-plane radius governs the propagation length within the polymer and therefore determines the MFD at the coupler's surface.

The optimization is carried out in an 8-dimensional geometrical parameter space. Results show, that a reduced set of 3+1 derived physical parameters describe the optimization problem in good approximation, which are MFD focus $w_f$, focal length $l_f$, propagation angle $\theta$ and to a lesser extent, MFD surface $w_s$.

As an empirical constraint, the gap between the fiber array and the coupler should be between 40-50 μm yielding a spot size at the couplers surface of $w_s$= 13-14 μm, see equation (1). This constraint results from a compromise between production time of the coupler and the robustness with respect to the variations of the z-alignment tolerances.

To link the coupler's physical properties to transmission efficiency, we simulate various designs using gridded parameter optimization, which we then fabricate and test on a $Si_3N_4$ chip with U-shaped backloops. A Scanning Electron Microscope (SEM) image of a fabricated backloop interfaced with TIR couplers is depicted in Fig. 2(b). Notably, losses are attributed solely to coupling, as on-chip propagation losses are negligible.

We determine the simulated transmission efficiency with the overlap integral at the coupler's focal point under an angle of 11.8 °, matching the evanescent beam's angle from the fiber array. Using an automated 3-axis stage for fiber array to PIC alignment, we measure experimental values and subtract Fresnel losses (0.15 dB) to align with theoretical predictions. We are subtracting the Fresnel losses for all the following experimental data as they are an inherit property of the air to polymer material transition.

A comparison of simulated and measured peak values at 1550 nm for varying coupler geometries is shown in Figs. 2c) and 2d). Each data point in Fig 2d) represents the optimized transmission for an individual coupler following positional alignment.

The experimental measurements align with the simulation's prediction that the evanescent beam's propagation direction is the primary factor influencing transmission efficiency. We attribute the 1° angle offset between simulated and fabricated couplers to slight shrinkage during fabrication. At optimal angles, simulation and experimental results differ by only -0.35 dB. However, deviations increase significantly to -3.5 dB leaving those optima, which we attribute to differences in data acquisition methods. In simulations, the overlap integral is calculated at the focal point of the coupler, whereas experimental transmission is measured by directing a light beam through one U-bend including two couplers at a fixed fiber array angle. The experimental result is then averaged over the two couplers. Since the emission angle of the coupler varies based on its physical parameters, while the fiber array consistently emits at 11.8, any angular mismatch results in an additional loss that is not accounted for in the simulations (Supplement A2).

In general, loss increases when the MFD at the coupler exceeds the fundamental mode of the SMF-28 fiber (10.4 µm), causing suboptimal power transfer between the fiber array and the coupler.

Deviation from the optimal power transfer angle causes a tilting of the mode profile relative to the fiber and effectively increasing the perceived MFD. Consequently, smaller mode field diameters achieve higher transmission efficiency leaving the optimum.

In the polymer, the MFD reaches 13–14 µm at the coupler, corresponding to a free-space propagation length of 40–50 µm to match the SMF-28 fiber's divergence. Deviations from this focal range reduce coupling efficiency.

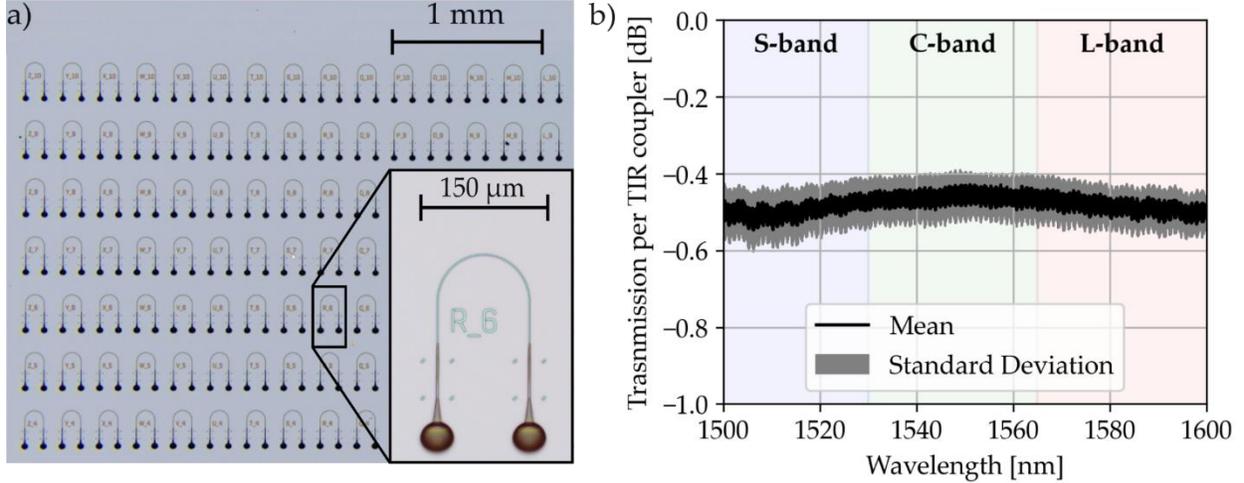

**Fig. 3** *TIR-coupler: Mass fabrication and Transmission Spectrum:* (a) Microscope image of a PIC featuring U-shaped Si$_3$N$_4$ waveguide backloops. The waveguides are interfaced with various TIR coupler geometries to determine the optimal coupling parameters. (b) Mean and standard deviation of the transmission spectrum of nominally identical TIR couplers, measured across the wavelength range of 1500 nm to 1600 nm, covering the S-, C-, and L-bands.

A microscope image of a test chip featuring multiple U-shaped backloops and couplers is presented in Fig. 3(a). The couplers are fabricated using two-photon grayscale lithography, allowing each coupler to be produced in under 1 minute. The average and standard deviation of the transmission spectrum measured from 1500 nm to 1600 nm across three nominally identical couplers are displayed in Fig. 3(b). The peak mean transmission reaches -0.41 dB, with the highest recorded transmission for an individual coupler reaching -0.39 dB.
Furthermore, the mean coupling efficiency across the S-, C-, and L-bands deviates by only -0.15 dB, indicating nearly wavelength-independent, high coupling performance within this range. Additionally, the low average standard deviation of -0.05 dB across the spectra demonstrates the reproducibility and consistency of the coupler fabrication process.

*2.2 Plug-and-Play Solution*

We are using the presented nearly wavelength-independent, ultra-low-loss coupler in a macroscopic plug-and-play solution for interfacing PICs. A female MTP cable, a fiber array with alignment pin holes is combined with alignment pins directly 3D printed using TPP on top of the PICs surface . By plugging the MTP cable onto those pins, the couplers are passively aligned with the PIC in a removable and repeatable manner.

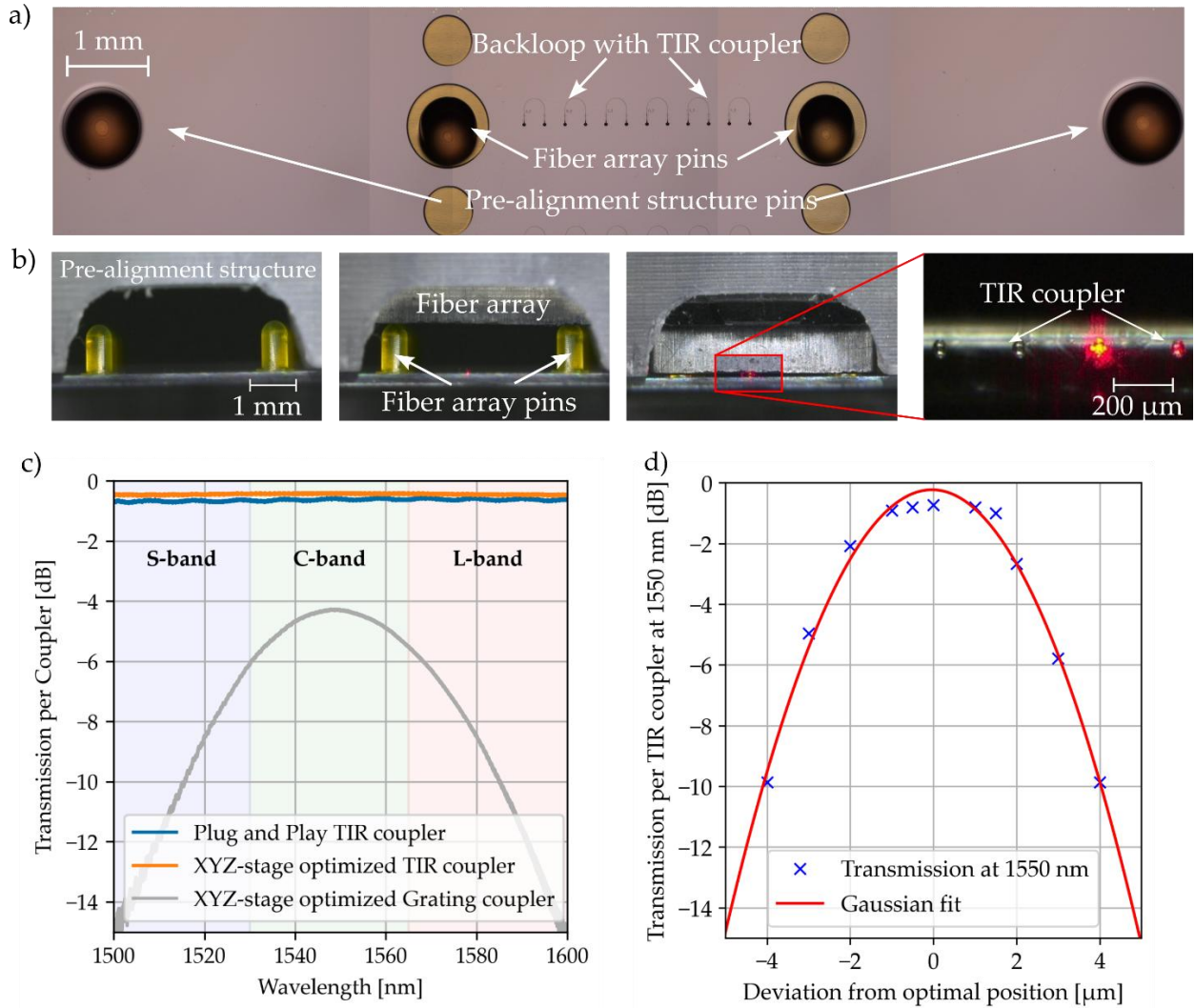

**Fig. 4** *Plug-and-play solution: Realization and tolerances:* (a) Microscope image of a Si$_3$N$_4$ chip demonstrating a plug-and-play solution for PICs. The MTP cable is passively aligned to the PIC using 8° angled fiber array pins, while pre-alignment pins enable exact positioning of the pre-alignment structure relative to the PIC. (b) A side-view microscope image shows the MTP cable alignment process via the pre-alignment structure and pins. Proper alignment is confirmed by the flickering of the TIR coupler under a red alignment laser. (c) Comparison of the transmission spectrum in between 1500 nm and 1600 nm of the plug-and-play solution with an XYZ-stage optimized TIR coupler and grating coupler. (d) Alignment tolerances parallel to the surface of the chip for the plug-and-play solution characterized by transmission at 1550 nm versus intentional offsets.

A microscope image of the passive alignment structure for the interfacing of a PIC comprising U-shaped backloops is shown in Fig. 4(a). The fabrication process involves two TPP steps: In the first TPP step, we use a high resolution 63x magnification objective to print the TIR couplers aligned with the $Si_3N_4$ waveguide tapers. In the second TPP step, we utilize a 10× objective to write the passive alignment structure, taking advantage of its large write field and high working distance to create millimeter-high structures.

The 8° angled alignment pins allow the polished facet of the MTP cable to be parallel to the chips surface, enabling access to a 2D grid of ports on the PIC. Spacers prevent the fiber array from crashing into the PIC and position the MTP cable at the couplers' focal point. The DLP printed pre-alignment structure aligns with the PIC through the same attachment mechanism as the MTP cable, a second set of TPP-printed pins. This improves system stability by supporting the MTP cable and dampening external vibrations. Figure 4(b) shows a side view of the MTP cable as it is plugged onto the PIC, with alignment confirmed by the TIR coupler flickering under red laser illumination.

Using the presented combination of pins and TIR coupler, we achieve a peak transmission of -0.55 dB. The transmission spectrum between 1500 and 1600 nm of the plug-and-play solution is compared to the one of the employed TIR coupler and a state-of the art grating coupler after 3 axis stage optimization in Fig. 4(c). The plug-and-play solution demonstrates only a -0.37 dB average difference in transmission compared to the optimized XYZ-stage setup, with significantly higher peak transmission and improved wavelength independence compared to traditional grating couplers. The 0.37 dB additional losses result from misalignment in between the TIR couplers and the fiber.

We evaluate the impact of MTP cable misalignment relative to the chip surface by intentionally offsetting the alignment loops from the optimal fiber-to-coupler position. As shown in Fig. 4(d), the transmission at 1550 nm follows a Gaussian distribution, with a -1 dB alignment tolerance of 4 μm, consistent with the findings of [25].Within a ±1.5 μm range from the optimal position, misalignment-induced loss remains below 0.3 dB—well within the 1.5 μm precision tolerance of the 10x objective used for fabricating the plug-and-play infrastructure.

## 2.3 Plug-and-play photonic computing

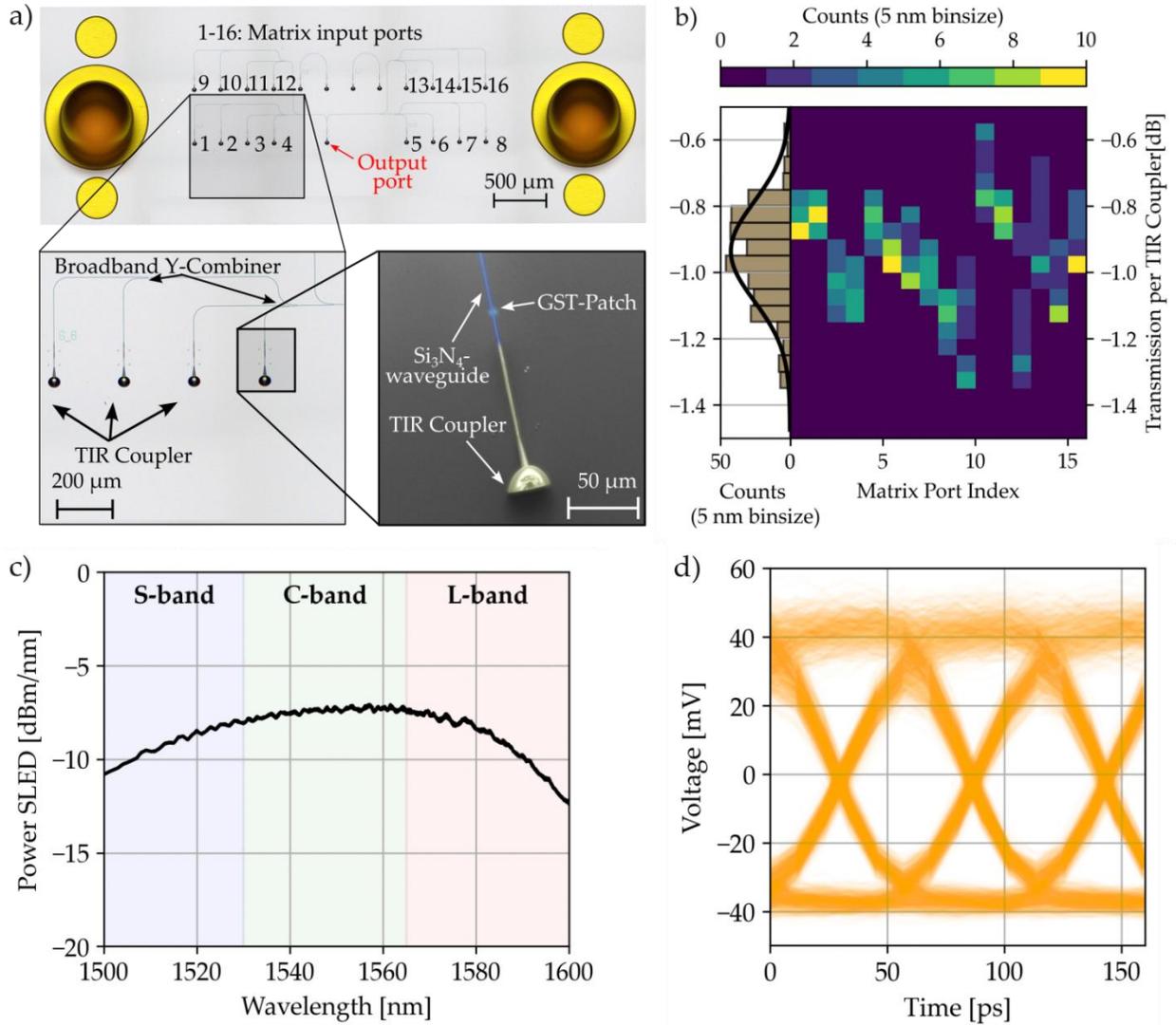

**Fig. 5** *Multiport interfacing of a 16x1 photonic matrix:* (a) Microscope image of a 16x1 photonic matrix interfaced with TIR couplers and addressable using the presented plug-and-play solution. The left inlet shows a close-up of 4 matrix input ports, whereas the left inlet is showing a SEM image of a waveguide loaded with the phase change material GST representing the matrix weights. (b) Colormap of the coupler transmission between 1500 nm and 1600 nm with 5 nm binsize, resolved for the different matrix ports. The histogram displays the transmission behavior across all matrix ports. (c) Transmission spectrum in between 1500 nm and 1600 nm of the SLED as broadband lights source for chaotic computing. (d) Eye diagram of the modulated SLED light after passing two couplers.

The demonstrated plug-and-play solution, combined with a 12×2 MTP cable, seamlessly interfaces with a 16×1 photonic matrix, as shown in the microscope image in Fig. 5(a). The matrix supports multiply-and-accumulate (MAC) operations, utilizing the MTP cable ports as input and output vectors and the phase-change material germanium-antimony-tellurium (GST), integrated on top of photonic waveguides as the matrix weights, shown in the SEM-image of Fig. 5a. Broadband Y-combiners are used to route light from the 16 input ports to the output port, with each GST-loaded input port connected through four Y-combiners.

The transmission behavior across the 16 channels is illustrated in Fig. 5(b). The colormap depicts the coupler transmission over a wavelength range of 1500 nm to 1600 nm, resolved in 5 nm bins for different matrix ports, while the histogram presents the combined transmission of all 16 ports. The results demonstrate the excellent reproducibility of the plug-and-play solution, with an average minimal coupler transmission of -0.78 dB and nearly wavelength independent coupling across the wavelength range. Moreover, they confirm the position independence of the coupling and highlight the broadband performance of the solution, with an average variation of only 0.36 dB between 1500 nm and 1600 nm. The high reproducibility over multiple attachment/detachment cycles, with a standard deviation of 0.14 dB, shown in Supplement A.3, underscoring the solution's consistency across both different ports and repeated connection cycles.

We use the TIR couplers in combination with the chaotic emission of a Super luminescent Light Emitting Diode (SLED) shown in Fig. 5(c). Using such broadband light sources enables a wide scale of applications, which benefit from using frequency multiplexing. For instance, large scale photonic crossbar arrays for matrix vector multiplication employ different optical carriers using the same frequency range [41]. Furthermore, chaotic light allows truly random photonic number generation, which is crucial for uncertainty estimation in probabilistic computing[42]. For both applications the available optical bandwidth is crucial to minimize the noise for the "deterministic" MVMs and to maximize the number of parallel samplings for the "probabilistic" MVMs. We modulate the SLED emission at a symbol rate of 17.6 GBaud and measure the eye-diagram after passing two couplers, see Fig. 5(d), translating to 1 MVM-operation per 56.8 ps. The high bandwidth of the SLED and the polymer couplers strongly suppress the noise in comparison to measurements with limited optical bandwidth [41].

## 3 Discussion and Outlook

Low-loss and broadband fiber-to-chip coupling is a critical enabler for a wide range of applications, from sensing [1], [2], [3] to optical communication [7], [8] . In this work, we have demonstrated a detailed optimization of out-of-plane polymer couplers based on total internal reflection, with an in detail comparison of FEFD-simulations and experimental results This comparison revealed the key physical parameters influencing coupling efficiency and provided insights into similarities and differences between simulated and fabricated couplers. The results facilitate adaptation to other platforms such as silicon-on-insulator (SOI) [43], lithium niobate ($LiNbO_3$)-[44], [45] and tantalum pentoxide [46], [47], [48], where only the mode transfer section needs to be adapted or other wavelength regimes.

**Table 1:** *Comparison of passive PIC interfacing techniques.* Comparison of passive PIC interfacing techniques in terms of minimum overall loss [dB], including packaging loss [dB], -1 dB threshold, interfacing method, and packaging complexity. The variance of the transmission is specified for the cases examined. The best results are indicated in **bold**. For the Jimenez [36] approach, marked with a star, the packaged transmission is compared to the one after active alignment, resulting in positive packaging loss. For this reason, this approach was excluded in terms of packaging loss.

|  | Minimum loss [dB] | Of which packaging loss [dB] | -1 dB threshold | Interfacing | Packaging fabrication complexity |
|---|---|---|---|---|---|
| Wan [39] | 5.67 | 0.67 | <10 nm | Grating Couplers | ++ |
| Jimenez [36] | 4.75 | 0.05 ± 1* | <50 nm | Grating Couplers | ++ |
| Bundalo [38] | 4.1 | 1.5 | <30 nm | Grating Couplers | +++ |
| Barwicz [34], [49] | 1.3 | No distinction | >100nm | Facet/In plane | + |
| Scarcella [37] | 1.7 | 0.4 ± 0.2 | <40 nm | Grating Couplers | +++ |
| Nair [35] | Not demonstrated | 1.55 | Not demonstrated | Not demonstrated | ++ |
| This work | **0.78** | **0.37 ±0.11** | **>>100 nm** | Polymer TRC Couplers | + |

Polymer couplers based on total internal reflection [26], [30] exhibit higher transmission and greater robustness than those relying on continuous bending [25], [27], [28], reducing loss and fragility. Unlike prior approaches requiring inlets on the chip surface [30], our method achieves superior performance without such advanced fabrication steps.

The presented total internal reflection coupler deliver the highest absolute transmission of -0.41 dB, with maximum 0.15 dB of deviation across the 1500–1600 nm wavelength range spanning over the C-, L-, and S-band. The reproducibility is demonstrated by a low standard deviation of 0.05 dB for nominal identical couplers over the mentioned wavelength regime.

We further demonstrated the reliable, removable and passive packaging of these couplers using a female MTP cable and monolithically integrated alignment pins on the chip surface, incurring only 0.37 dB additional loss at a standard deviation of 0.11 dB for nominally identical devices at a wavelength of 1550nm.

Our approach offers superior performance compared to prior out of plane passive packaging solutions presented in table 1, where grating couplers have been packaged with losses between 5.67 and 1.7 dB [35], [36], [37], [38], [39] at a maximum -1dB threshold of 50 nm[36]. The overall performance of the presented solution is comparable to the one of in-plane coupling approaches using on-chip V-grooves [49] but with the benefit of a higher number of addressable ports.

Integrating alignment pins directly on the surface of the chip offers the design flexibility of out of plane coupling, eliminating the need for precise glueing of the chip relative to the PCB, as required by approaches where MTP alignment pins are integrated on the PCB [50] . Our solution accommodates a wide range of chip sizes, overcoming limitations imposed by the spacing between MTP cable ports. Additionally, the spacer printed on top of the PIC ensures that the fiber array is precisely aligned in the optimal horizontal position for coupling, providing optimized alignment across all three axes.

Alternative material platforms such as AlGaN [51] and half-etched $LiNbO_3$[52], [53] , the mode transfer to a polymer waveguide is impossible due to the high refractive index of the substrate. As we realize the passive packaging and the photonic interfacing in 2 different processes, the presented plug-and-play is applicable to all types of out-of-plane couplers, including all state of the art grating couplers, making it very versatile. Also, the low refractive index polymer (1.53) serves as a cladding for most on-chip waveguides, meaning that the area underneath the TPP-printed pillars remains available for additional photonic components, optimizing chip area use.

Using in-situ exchange of the photo resin during the printing process [54] can decrease the fabrication time by reducing the number of development steps. Also a higher resolution for the 10x objective could enable to merge the photonic and the packaging TPP step further reducing the fabrication time.

The presented plug-and-play solution, combined with the highly efficient broadband coupler, also enables chip-to-chip coupling for reconfigurable architectures and pre-testing of components. Overall, this marks a significant step towards reproducible, reconfigurable, and passive packaging, paving the way for scalable mass production of photonic hardware.

# 4 Methods

The devices are fabricated on a silicon wafer with a 330 nm silicon nitride ($Si_3N_4$) layer atop a 3.3 µm silicon dioxide ($SiO_2$) layer (Rogue Valley Microdevices). The fabrication process begins with the spin-coating of positive resist polymethylmethacrylate (PMMA) to define gold markers and pads. The mask is patterned using a 100 kV electron beam lithography (EBL) system (Raith EBPG5150), which is used for all subsequent lithographic steps. A 5 nm chromium (Cr) adhesion layer and an 80 nm gold (Au) layer are deposited via physical vapor deposition, followed by a lift-off process using acetone. Next, the negative resist AR-N 7520.12 is spin-coated and the mask for photonic structures is aligned to the previously defined gold markers and patterned. After developing the resist, the silicon nitride layer is fully etched using reactive ion etching (RIE) with $CHF_3/O_2$ plasma, and the resist is subsequently removed with oxygen plasma. A layer of PMMA is then spin-coated for the deposition of the phase change material (PCM) germanium-antimony-tellurium (GST), and the GST regions are exposed and developed. A 15 nm GST layer, capped with a 10 nm $Al_2O_3$ layer to prevent oxidation, is deposited using radio-frequency sputtering with argon plasma.

Following lift-off, three-dimensional polymer couplers are fabricated at designated locations using automated aligned Two Photon Polymerization (TPP) [55] on a Nanoscribe QX system with the 63x high resolution objective. The first TPP step employs grayscale two-photon lithography, enabling a tenfold increase in coupler fabrication speed compared to conventional two-photon lithography. After the coupler development in PGMEA, a second TPP step utilizes traditional two-photon lithography with the 10x objective to write the infrastructure necessary for interfacing the total reflection couplers with a fiber array, which is then developed in a PGMEA bath.

Digital light processing (DLP) 3D printing was performed on an Asiga MAX X27 UV DLP 3D printer with a LED light source of 385 nm and a pixel resolution of 27 μm. All parts were printed with a commercial bisphenol A (BPA)-based ink (moiinTechClear). The prints were performed at 30 °C with a layer thickness of 100 μm. The optimal exposure time was set to 2.1 s for an intensity of 10 mW cm$^{-2}$. After printing, the parts were removed carefully using a thin blade. To remove excess ink, the parts were washed and sonicated two times in isopropyl alcohol. The structures were dried with pressurized N2 and postcured in an Asiga UV chamber for 2 min.

The transmission spectrum from 1500 nm to 1600 nm is measured using a Santec TSL-710 tunable laser source. For the coupler optimization, a standard fiber array (SMF28) polished at an 8° angle is positioned slightly above the sample to align. The system's nanometer-precision xyz stages and closed-loop positioning enable automated optimization of chip-to-fiber coupling through orthogonal scans, ensuring efficient and accurate characterization of all devices.

For the measurements related to the plug-and-play approach a 12x /24x female MTP breakout cable was connected to the a Santec TSL-710 tunable laser source.

*Disclosures*

The authors declare that there are no financial interests, commercial affiliations, or other potential conflicts of interest that could have influenced the objectivity of this research or the writing of this paper.

*Code, Data, and Materials Availability*

The data that support the findings of this study are available from the corresponding author upon reasonable request.

*Acknowledgments*

E.J., W.P., E.B. and C.V.M. acknowledge the funding from the Deutsche Forschungsgemeinschaft (DFG, German Research Foundation) via the Excellence Cluster "3D Matter Made to Order" (EXC-2082/1-390761711) and the Carl Zeiss Foundation through the Carl-Zeiss-Foundation-Focus@HEiKA. C.V.M. acknowledges the Fonds der Chemischen Industrie for the support during her PhD studies through the Kekulé Fellowship. H.G. thanks the Studienstiftung des deutschen Volkes for financial support. We thank Shabnam Taheriniya for the help with the SEM images in Figure 2 and 5.

**First Author** is a PhD student at Heidelberg University. He received his BS and MS degrees in physics from KIT in 2019 and 2021. His current research interests include integrated photonics, direct laser writing, and photonic packaging.

**Caption List**

**Fig. 1** Plug-and-play concept.

**Fig. 2** Coupler Optimization.

**Fig. 3** TRC-coupler: Mass fabrication and Transmission Spectrum.

**Fig. 4** Plug-and-play solution: Realization and tolerances.

**Fig 5** Multiport interfacing of a 16x1 photonic matrix.

**Table 1:** Comparison of passive PIC interfacing techniques.